# Simulation on the Miniaturization and Performance Improvement Study of Gr/MoS$_2$ Based Vertical Field Effect Transistor

*Sirsendu Ghosh[1], Anamika Devi Laishram[2], and Pramod Kumar[1]\**

[1]Department of Physics, Indian Institute of Technology Bombay, Mumbai, Maharashtra, 400076, India

[2]Center for Research in Nanotechnology and Science (CRNTS), Indian Institute of Technology Bombay, Mumbai, Maharashtra, 400076, India

E-mail: pramod_k@iitb.ac.in

**Abstract**

Vertical field effect transistors (VFETs) show many advantages such as high switching speed, low operating voltage, low power consumption, and miniaturization over lateral FETs. However, VFET still faces the main challenges of high off-state current. Graphene (Gr) and transition metal di-chalcogenides (TMDs) are attractive materials for the next generation electronics. In this simulation work, the bulk molybdenum disulfide (MoS$_2$) is sandwiched between perforated monolayer Gr which acts as the source electrode, and aluminum (Al) as the top drain electrode. In addition to this, the minimization of the off-state current is carried out by modifications in the source contact geometry by insulating some part of the source electrode and introducing the extra MoS$_2$ layer between the source and gate dielectric named as buried layer. After the modification, the results show an improvement in OFF current, hence the ON/OFF ratio. The highest ON/OFF ratio of $10^9$ is achieved with top side insulated source contact and thinnest buried layer of 02 nm with top and sidewall insulation. These results would support low voltage operation with high switching speed in complete 2D material based VFETs and further miniaturize its geometry.

**Keywords**: Vertical field effect transistors (VFETs), 2D materials, Graphene, MoS$_2$, Simulations

## 1. Introduction

Research and development in the field of vertical field effect transistors (VFETs) shows that it can provide high current density over the planer field effect transistors (FETs) [1]. In VFET geometry, the channel is sandwiched between the source and drain electrodes, unlike the planer FET structure. The charge carrier transport takes place vertically from source to drain. The gate field interaction occurs through the perforation on the bottom source electrode, which plays a major role in switching the device. Gate field penetration from dielectric requires some area directly in touch with the semiconductor, hence perforations are made on the bottom source electrode [2]. The gate field penetrates and changes the source-channel Schottky barrier height (SBH) to inject or stop the charge carrier injection from the source to the semiconductor. Graphene (Gr) is one of the most emerging materials over the past decade with outstanding properties like 2D nature, high charge carrier mobility, tunable work function, and low density of states [3], [4], [5]. The gate field can easily modulate the Gr-channel SBH due to the low density of states in it, hence Gr is now desirable in the scientific community as a source electrode in VFETs [6], [7], [8], [9]. However, a source electrode based on a continuous Gr single layer or a few layers degrades the device's performance due to the high electrostatic screening effect of the gate field. Perforations on the Gr layer can help in gate field penetration and get the high ON/OFF ratio in low voltage operation VFET [10], [11], [12]. The source electrode perforation size and its thickness have shown a great effect on the performance of organic semiconductor based VFETs, where the lowest thickness source electrode is conducive to better performance [13]. Gr is an excellent source contact to reduce the source electrode thickness below the nanometer range. It is shown that other optimizations can further improve the performance of the organic semiconductor based VFETs in terms of ON/OFF ratio and SS by introducing a top insulating layer on the source electrode and a thin buried semiconductor layer [14]. The drawback of organic semiconductor polymers that can be solution processed is that the continuous buried layer below 05 nm is experimentally difficult to achieve. Transition metal di-chalcogenides (TMDs) are promising materials as they possess semiconducting properties and hence can be used for low dimensional field effect transistors [15], [16], [17]. The advantage of TMDs is their two dimensional (2D) atomically thin semiconductor layers similar to Gr and therefore can offer solutions in further miniaturization and better-performing VFETs. TMDs can be easily exfoliated from bulk material or grown directly on the substrate and can offer monolayer thickness typically in the subnanometer range [18], [19], [20], [21], [22], [23]. TMDs have been one of the most important topics in science over the past few years due to their tremendous electrical, optical, and structural properties [24], [25], [26], [27]. Molybdenum disulfide ($MoS_2$) is one of the TMDs that is important due to its various applications in the fields of electronics, energy, biomedical, and sensing [28], [29], [30], [31], [32]. Various research groups have tried to explain the bulk as well as monolayer $MoS_2$ properties using many scientific experimental techniques like transient microscopy study, femtosecond laser irradiation, etc. [33], [34], [35], [36], [37], [38]. Among all the TMDs, $MoS_2$ field effect transistors (FETs) are of great importance in electronics due to their high charge carrier mobility and ON/OFF ratio [39], [40], [41], [42]. Hence $MoS_2$ can be useful in the case of VFETs as it can provide further miniaturization of the VFET structure and can remove some of the bottlenecks in the case of organic semiconductor based VFETs. In this paper, we discussed the parameters important to enhance the performance of $MoS_2$ based VFETs using a 2D simulation model in COMSOL Multiphysics. The device structure consists of a highly doped gate ($n^{++}$-Si), $SiO_2$ as a gate dielectric, source (monolayer Gr), $MoS_2$ (semiconductor), and Al as a top drain electrode. We varied the gate width (perforation on the Gr source contact), and channel length to get the optimized dimension for the best performance of the device. The high off-state current issue is addressed with source electrode modification by adding an extra insulator layer on the source electrode (Gr). Simulations were also performed by introducing the buried layer of $MoS_2$ of different thicknesses which led to significantly better performance in terms of both ON/OFF ratios and subthreshold swings (SSs).

## 2. Simulation details

COMSOL Multiphysics software is used for the simulation of the VFETs by two-dimensional model and using the semiconductor module. This module consists of the drift-diffusion theory of charge carrier transport. The details of the drift-diffusion theory are given elsewhere and in the supporting information [14]. In this simulation, $MoS_2$ is taken as the semiconductor channel, pyrolytic-graphite of monolayer thickness 0.34 nm for monolayer Gr [43], is taken as the perforated source electrode, and Al as the top drain electrode. The Gr source lateral width is taken as 10 nm for monolayer Gr throughout the simulations since below 10 nm has a finite band gap [44]. The source electrode makes Schottky type contact with the $MoS_2$ channel whereas the top drain electrode behaves as an ohmic contact. Finite-volume discretization of triangular-type mesh is considered throughout this work. Neumann boundary conditions are applied to the thin insulator gate. The parameters used in this simulation for $MoS_2$ are listed in Table 1:

Table 1. Basic properties of $MoS_2$ [45],[46], [47]

| Relative permittivity, $\varepsilon_r$ | Band gap, $E_{g0}$ [eV] | Electron affinity, $\chi_0$ [eV] | Effective density of states, valance band, $N_v$ [cm$^{-3}$] | Effective density of states, conduction band, $N_c$ [cm$^{-3}$] | Electron mobility, $\mu_n$ [cm$^2$ V$^{-1}$ s$^{-1}$] | Hole mobility, $\mu_p$ [cm$^2$ V$^{-1}$ s$^{-1}$] |
|---|---|---|---|---|---|---|
| 7.1 | 1.3 | 4.45 | $1.8 \times 10^{19}$ | $2.2 \times 10^{18}$ | 100 | 25 |

## 3. Results and discussion

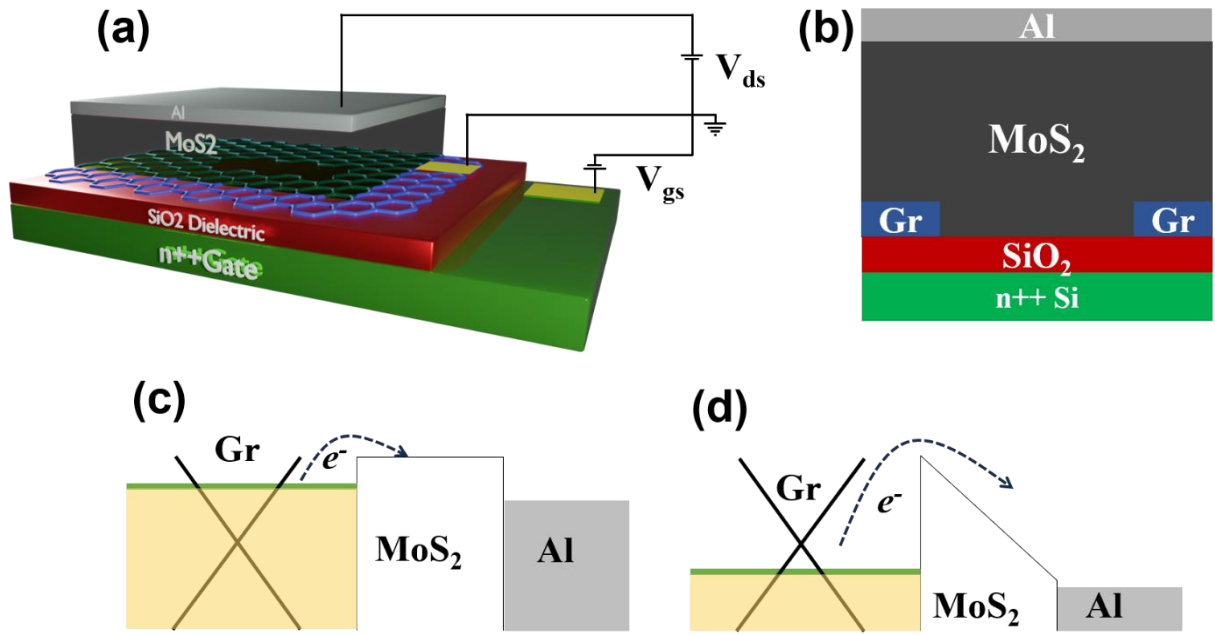

Figure 1. (a) 3D architecture of the Gr/MoS2 VFET device. (b) Corresponding single perforation 2D device geometry. Band diagram at Gr/MoS$_2$ interface at constant positive drain voltage (c) V$_g$ is positive (d) V$_g$ is negative.

The schematic diagram of VFET based on Gr and MoS$_2$ as the semiconductor is shown in Fig. 1 with the effect of the gate field on the SBH. Fig. 1 (a) and (b) respectively show the 3D schematics of the VFET and 2D cross-section of a single perforation on Gr source contact. The perforation on the Gr sheet helps in the gate field penetration which changes the SBH as shown in Fig. 1 (c) and (d) for positive and negative gate voltage (Vg), respectively. Simulations focused on a single perforation on the Gr source contact constituting the device's single perforation were carried out. The first stage optimization was done by varying the channel length/semiconductor thickness with fixed gate opening/source contact perforation of 50 nm, source height of 0.34 nm (Gr thickness), and source width of 10 nm (Gr width). The transfer characteristics, ON/OFF ratios, and subthreshold swings (SSs) are summarized in Fig. 2. The simulated transfer characteristics (J-V$_g$) show a rise in the off-state current with lower channel length as shown in Fig. 2 (a). The device performance improvement is noticed in terms of the extracted ON/OFF ratio and SS with the increasing channel length, as shown in Fig. 2 (b) and (c). The drain current modulation is less with a shorter channel length. The moderate ON/OFF ratio and SS value of $10^4$ and 508 mV/decade are achieved with a channel length of 100 nm. At positive drain voltage, the electron will go from source to channel. Consequently, the SBH between the source and channel is important here as shown in Fig. 1 (c) and (d). Now, the SBH increases at negative gate voltage leading to less current, considered as off-state current. At positive gate voltage, the SBH decreases. Therefore, the electrons can easily overcome the SBH to go to the channel known as on-state current [6]. The main reason behind the performance degradation on decreasing the semiconductor thickness is the rise in the off-state current, here the drain field dominates due to proximity with the source contact as shown in Fig. 2 (d) [1], [13]. The gate field penetration in the channel is limited but on increasing the thickness, the gate field has shown better penetration between the source and drain contact as shown by the yellow region in Fig. 2 (e), causing a decrease in the off-state current and sharp rise in current with gate field. Consequently, the ON/OFF ratio increases and SS decreases with an increase in $MoS_2$ channel thickness.

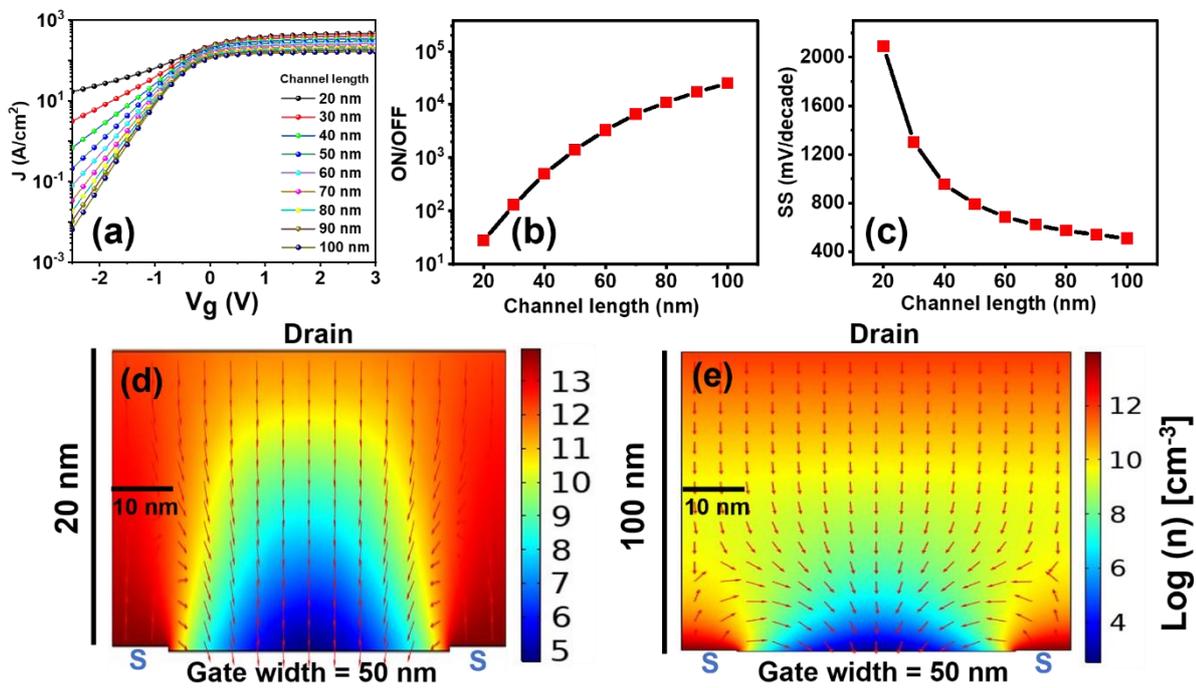

Figure 2. (a) Transfer characteristics of the device with the variation of the channel length at constant drain voltage $V_d$ = 0.1 V. (b) ON/OFF ratio and (c) SS of the device calculated from the transfer characteristics. Electron concentration profile (n) at $V_g$ = -2.5 V and channel length (d) 20 nm and (e) 90 nm, respectively.

In the second stage, the channel length is kept fixed at 100 nm and the gate width/perforation size is investigated for changes in device performance. The perforation sizes were varied from 10 to 100 nm and the transfer characteristics, ON/OFF ratios, and SSs are shown in Fig. 3. The transfer characteristics show improvement in terms of off-state current and the current rising slope as seen in Fig. 3(a) with increasing gate width (Gr perforation size). The ON/OFF ratios and SSs extracted from the transfer characteristics are shown in Fig. 3 (b and c) which show the best performance when the gate width is 100 nm.

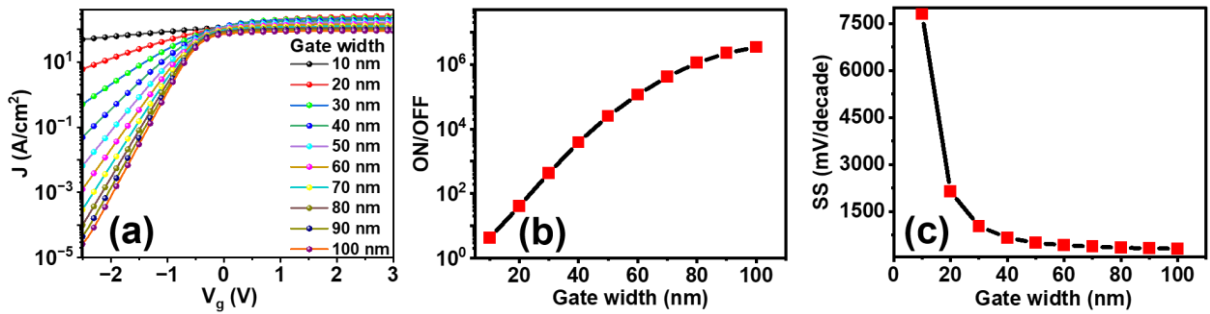

Figure 3. (a) Transfer characteristics of the device by varying the gate width at constant drain voltage $V_d = 0.1$ V. (b) ON/OFF ratio and (c) SS of the device calculated from the transfer characteristics.

By changing the gate width from 10 to 100 nm, the decrease in the off-state current improves the ON/OFF ratio which signifies the greater gate field affecting the current flow. The highest ON/OFF ratio of greater than $10^6$ and moderate SS of 316 mV/decade is achieved at a drain voltage of 0.1 V showing improvement over experimental results on fabricated devices without Gr as a source electrode by considering all the factors including ON/OFF ratio, SS and low power consumption [48], [49]. The greater opening area on the source electrode provides a higher effect of the negative gate field causing less off-state current. The negative gate field will play efficiently with more gate perforation leading to higher SBH for electrons to flow from source to channel. Hence, off-state current decreases with increasing gate width. In the third stage of optimization to further reduce the off-state current an insulating layer is placed on top of the Gr source electrode by keeping the gate perforation at an unoptimized value of 50 nm and channel length/semiconductor thickness is varied from 20 to 100 nm. In this case, the top side of the source electrode is insulated, hence the off-state current can be further reduced. The device transfer characteristics, ON/OFF ratios, and SSs are shown in Fig. 4. Significant lowering of the off-state current is observed as shown in Fig. 4 (a) if compared without insulation in Fig. 3 (a). The ON/OFF ratios and SSs show improvement with the insulating layer on Gr source contact as shown in Fig. 4 (b and c). Only the side walls are responsible for the injection/extraction of the charge carrier with the insulation on the top side of the Gr source contact. The comparison of with and without insulation of the top are shown in Fig. 4 (d) and (e), respectively. The highest ON/OFF ratio of $10^8$ is achieved with a channel length of 100 nm. However, it should be noted that here Gr perforation/gate width is only 50 nm which produces the same ON/OFF of $10^6$ as we observed in the case of without insulation on Gr source contact at 100 nm channel length and 100 nm gate width. So, the insulation on top of the Gr source contact can reduce the device size by half which is directed toward the miniaturization of the device structure.

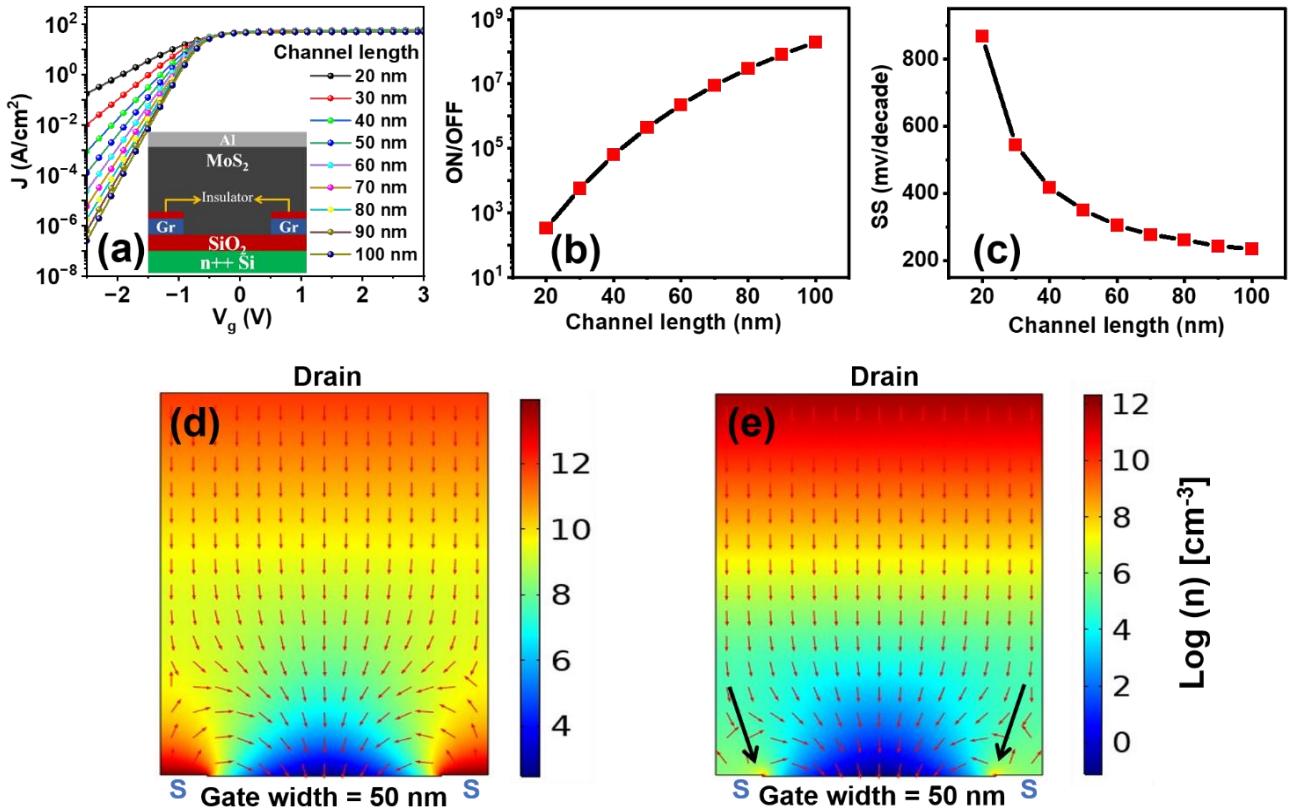

Figure 4. (a) Transfer characteristics of the device with channel length variation having source top side insulation at constant drain voltage $V_d$ = 0.1 V. The inset shows the device geometry where the top side of the Gr source is insulated. (b) ON/OFF ratio and (c) SS of the device calculated from the transfer characteristics. Electron concentration profile (n) at $V_g$ = -2.5 V of 100 nm channel length (d) no insulation at source and (e) top side of the source contact is insulated, respectively.

The gate width can also be investigated for the miniaturization of the device to find the effect of insulation on top of the Gr source contact. The gate width is varied from 10 to 100 nm and the results are summarized in Fig. 5. Fig. 5 (a) shows the transfer characteristics at different gate widths which clearly show a reduction in the gate field effect at lower gate widths as the gate field is not able to penetrate in the semiconductor. The ON/OFF ratios and SSs show enhancement with gate width initially and getting almost saturated at around 60 nm gate width as shown in Fig. 5 (b) and (c). The best device performance having an ON/OFF ratio of $10^9$ and SS of 193 mV/decade is found for the gate width of 100 nm. Overall, the results show that by considering the top side insulation, the device having 30 nm gate width has shown the same order of ON/OFF ratio $10^6$ as the gate width of 100 nm with no insulation on top. Hence, miniaturization of the gate width is possible with the insulating layer on the Gr source contact.

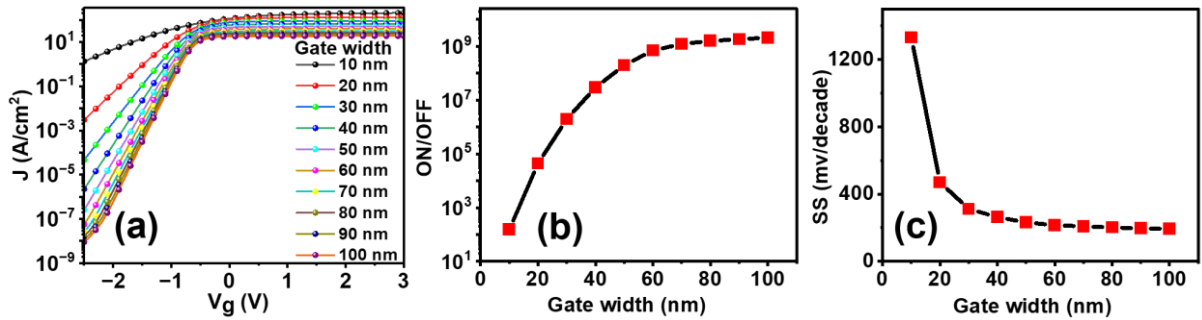

Figure 5. (a) Transfer characteristics of the device with gate width variation having source top side insulation at constant drain voltage $V_d = 0.1$ V. (b) ON/OFF ratio and (c) SS of the device calculated from the transfer characteristics.

A buried $MoS_2$ semiconductor layer is introduced below the perforated Gr source contact to further optimize the device's performance. The extra semiconductor layer between the source electrode and gate dielectric is defined as a buried layer and has shown better control of the gate field on the SBH [14]. There is no direct contact between the source and gate dielectric in this structure. Starting from a buried layer thickness of 20 nm keeping gate width and channel length fixed at 100 nm, the buried layer thickness is decreased to see its effect on the device performance. The buried layer geometry can be also divided into three different conditions based on the location or existence of the insulating layer on the Gr source electrode. Simulations were carried out for these different conditions like "without insulation", "insulation on the top side", and "insulation on top and sidewall of the Gr source contact". The device geometries and the summary of the results are shown in Fig. 6. The three device schematics are shown in Fig. 6 (a) and are marked as case 1, case 2, and case 3. The transfer characteristics show lowest off-state current and highest current rising slope in the case of insulated top and side walls of the Gr source contact as shown in Fig. 6 (b). Approximately three orders magnitude higher ON/OFF ratio is achieved in case 3 where insulation is applied on top and sidewall of the source contact. Overall the best performance has been shown by the device when the top and sidewalls are insulated having an ON/OFF ratio of greater than $10^6$ and SS of 290 mV/decade as shown in Fig. 6 (b). By insulating the top side of the source contact, one possible path is removed to provide the off-state current compared with no insulation. Similarly, by insulating the top and sidewall of the source contact, there is only one path to provide the off-state current. The bottom side of the Gr source electrode is not simulated with an insulating layer as it directly faces the gate field and hence the effect of SBH variation is much higher [14]. The off-state current and current rising slope show improvement in this condition due to the much larger effect of the gate field on the direct-facing bottom Gr source contact side. The buried layer thickness is also varied at this optimization condition by decreasing the buried layer thickness to 10 nm and 02 nm, and are shown in Fig. 7. Fig. 7 (a) shows the transfer characteristics at three different buried layer thicknesses showing the greater effect of the gate field on SBH with lowering buried layer thickness giving rise to least off-state current and highest current rising slope. Fig. 7 (b) compares the ON/OFF ratios and SSs at these buried layer thicknesses clearly showing further improvement in the device performance. The highest ON/OFF ratio of $10^9$ and SS value of 197 mV/decade is achieved with 02 nm buried layer thickness as shown in Fig. 7 (b). The buried layer geometry having a 02 nm semiconductor under Gr source contact with insulation on top and side walls with 40 nm channel can achieve the same order of ON/OFF ratio $10^6$ as the one without buried layer (no insulation) having a 100 nm channel thickness. In fact, the case of a very thin buried layer with a gate width of 30 nm (channel length 100 nm) shows an ON/OFF ratio of $10^6$ similar to the one without a buried layer and insulation at a gate width of 100 nm. Hence, introducing a buried layer with Gr source contact's top and sidewall insulation leads to further miniaturization of the device, since $MoS_2$ monolayer thickness is in subnanometer range.

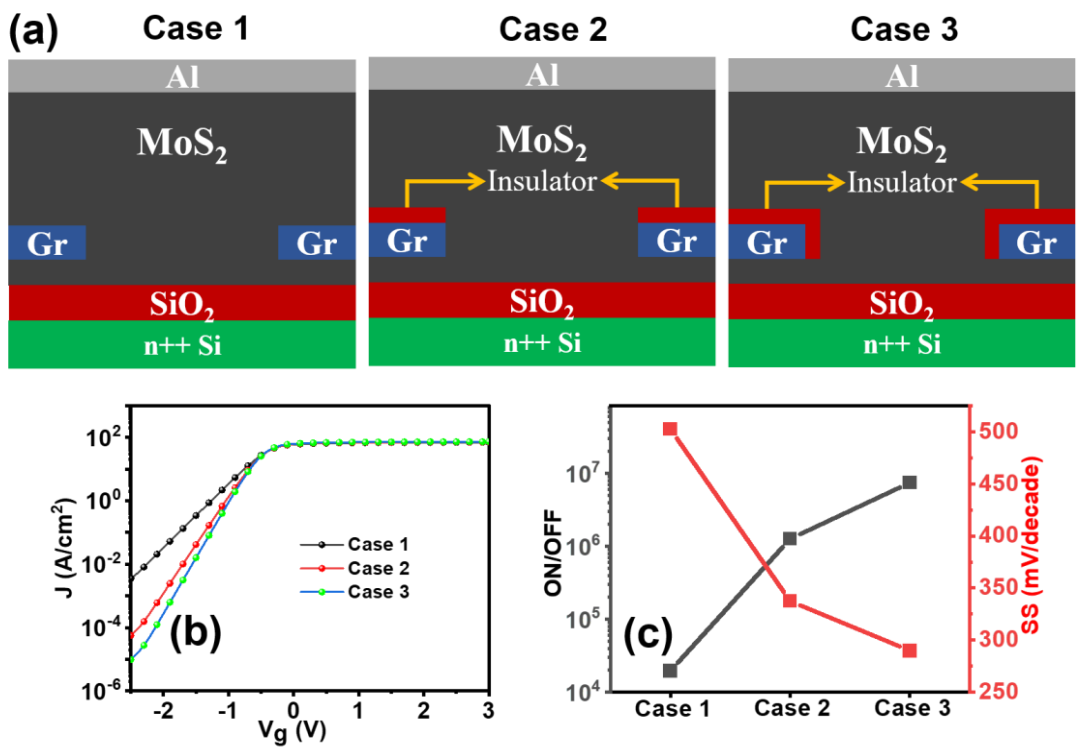

Figure 6. (a) Schematics of 2D device geometry having buried layer thickness 20 nm in three cases. Case 1: No insulation in the source contact, Case 2: Top side, and Case 3: Top and sidewall of the source contact, are insulated respectively. (b) Transfer characteristics of the device of all three cases at constant drain voltage $V_d$ = 0.1 V. (c) ON/OFF ratio and SS of the device calculated from the transfer characteristics.

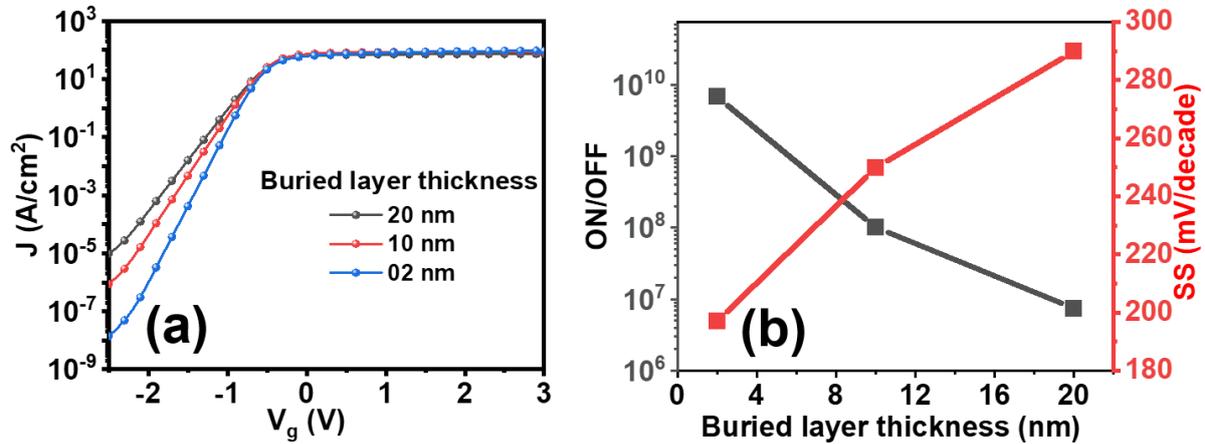

Figure 7. (a) Transfer characteristics of the device with varying buried layer thickness at constant drain voltage $V_d$ = 0.1 V. (b) ON/OFF ratio and SS of the device calculated from the transfer characteristics.

## 4. Conclusions

In this article, we compared the transfer characteristics of Gr/MoS$_2$ based VFET by modifying the Gr source electrode and the thickness of the MoS$_2$ layers with COMSOL Multiphysics software. First VFET is optimized for both MoS$_2$ channel length/thickness and Gr source contact perforation to get the best-performing device. In the second stage of optimization, the unwanted high off-state current is reduced by introducing an insulating layer on top of the Gr source electrode. This insulating layer on the source electrode improved the performance and also miniaturized the VFET device by half. The highest ON/OFF ratio of $10^9$ and lowest SS of 193 mV/decade is achieved in this case. The final modification of the Gr/MoS$_2$ based VFET is done by introducing a buried layer of MoS$_2$ below the Gr source contact and adding insulating layers on top and side walls of the source contact. The results show that by decreasing the buried layer thickness with top and sidewall insulation improved the ON/OFF ratio and SS. Finally, the VFET without buried MoS$_2$ layer with an insulating layer on the top of the Gr source contact and the VFET having a buried layer of 02 nm with top and sidewall insulation on Gr source contact showed the best performance having ON/OFF ratio in the order of $10^9$ and SS ~190 mV/decade. MoS$_2$ monolayers can further enhance the VFET performance due to its thickness in a few hundred angstroms. The results also show that the VFET size can be reduced by less than 50% in the modified Gr source contact with insulation. The miniaturization possibility can be further improved by a buried $MoS_2$ layer of subnanometer thickness below the Gr source contact. Therefore, miniaturization in terms of both channel length and gate width is possible without any compromise on the ON/OFF ratio and SS in both the geometry viz. with and without buried $MoS_2$ layer and with insulation on Gr source contact. Our results would give promising future direction for optimizing the device structure to get a high switching speed of Gr/TMDs based vertical transistors as TMDs can provide single-layer processing and fabrication of VFETs.

## Supplementary material

The supplementary material contains the simulation details of the Gr/MoS2-based vertical field effect transistor (VFET).

## Declaration of competing interest

We declare that we have no significant competing financial, professional, or personal interests that might have influenced the performance or presentation of the work described in this manuscript.

## Acknowledgements

The authors would like to thank the Indian Institute of Technology Bombay (IITB) for the funding and fellowship.

## Data availability

The data that support the findings of this study are available from the corresponding author upon reasonable request.